\documentclass{elsart}
\usepackage[english]{babel}

\usepackage{epsfig}
\usepackage{graphics}
\usepackage{cite}

\usepackage{amssymb}

\begin{document}

\begin{frontmatter}

\title{Volume Free Electron Laser - Self-Phase-Locking System}

\author{V.G. Baryshevsky and A.A Gurinovich}

\address{Research Institute for Nuclear Problems, Belarusian State
University, 11~Bobruiskaya Str., Minsk 220030, Belarus}
\ead{bar@inp.bsu.by, v\_baryshevsky@yahoo.com}

%\affiliation{Research Institute for Nuclear Problems}
\begin{abstract}
It is shown that that the Volume free electron laser is a
self-phase-locking system. Equations are derived that describe
both the phase-locking process and the dependence of generation on
the times of bunches' entering into the VFEL resonator in the
superradiation regime and for long pulses. The derived equations
are applicable to the description of the stated processes in
gratings formed by relativistic BWOs.
\end{abstract}

\end{frontmatter}

\section{Introduction}
The research into pulse power amplification in microwave
generators is gaining importance nowadays \cite{1,giri}. It has
been shown that even for short microwave pulses, the discharge
processes in the resonators of generators place limitations on the
radiation power amplification potential.  The possibilities of
coherent combining of fields generated by several microwave
generators  are actively studied in this connection as the means
of tackling these limitations. Particularly, phase-locking is a
subject of vigorous study \cite{giri,woo,whar,elch}.

It has also been demonstrated that two separate super-radiant
backward wave oscillators (BWO) connected to one and the same
voltage (power) supply of subnanosecond rise time can form two
 coherent waves, which causes a fourfold increase in the radiation power \cite{elch}.

 Thus, these experiments confirm that using $N$ number of BWOs in this way, two (or more)-channel
nanosecond relativistic microwave generators can be developed,
whose total power will be as high as $W\sim N^2$ ($N$ is the
number of BWOs \cite{whar}.

  Study of the application
potential of volume free electron lasers (VFEL) for the
development of relativistic microwave and optical generators with
increased power is another promising line of investigation in this
field \cite{bar12}.

\begin{figure}[htbp]
\begin{center}
\epsfxsize = 10 cm \centerline{\epsfbox{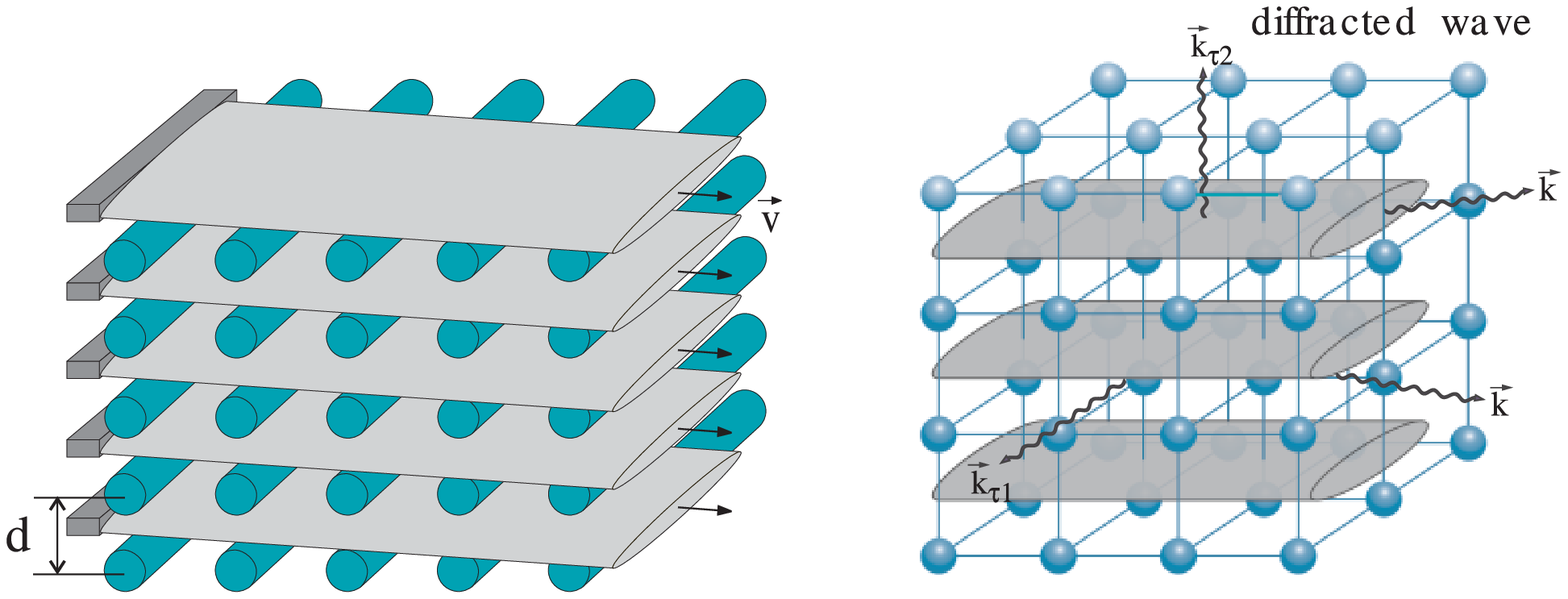}} \caption{}
\label{volume}
\end{center}
\end{figure}

\begin{figure}[h]
\begin{center}
%---\epsfxsize = 13 cm \
%---centerline{\epsfbox{multiwave.eps}}
\includegraphics[scale=0.6]{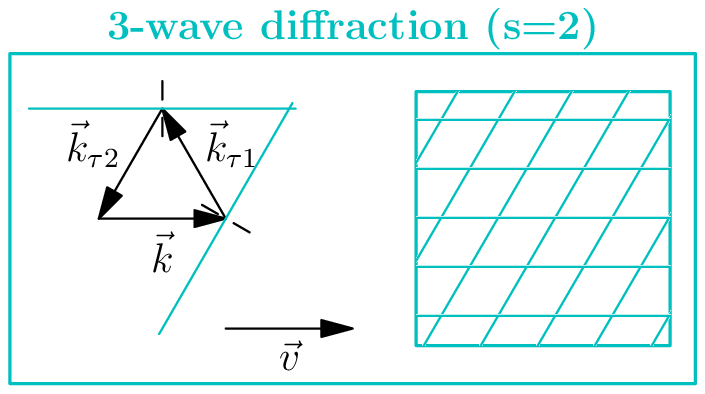}
\quad \includegraphics[scale=0.6]{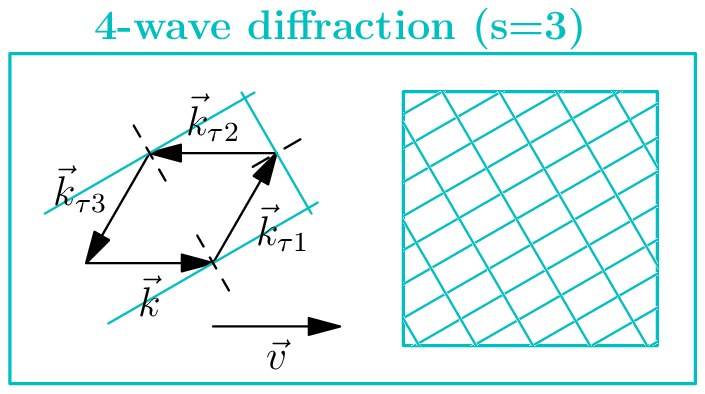}
\quad\includegraphics[scale=0.6]{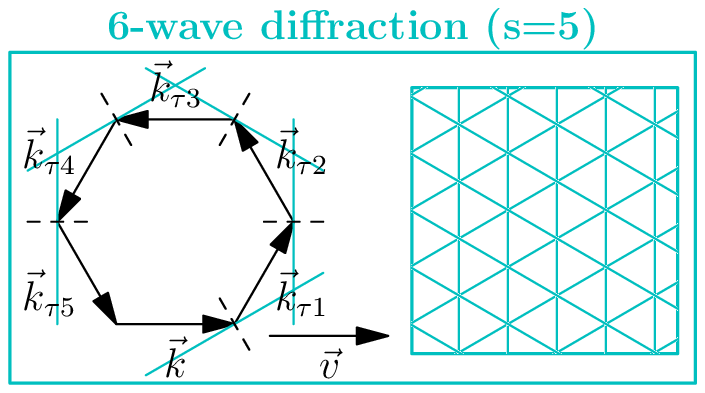}
\end{center}
\caption{} \label{multiwave}
\end{figure}

Two(three)-dimensional distributed feedback (DFB) arising in VFEL
resonators with two-three dimensional spatially periodic
structures (now often called electromagnetic or photonic crystals)
enables producing coherent radiation from wide electron beams or
several beams (see Figs. \ref{volume}, \ref{multiwave}). A key
feature of the VFEL is that as a result of diffraction, the signal
is transferred from one point to another, thus linking the points
of the beam and making them generate coherently (see Figs.
\ref{volume}, \ref{multiwave}).

This means that the VFEL is a
self-phase-locking generator of radiation (phase-locking inside).
When $N$ number of electron beams pass through electromagnetic
crystals, the radiated power
 will increase in proportion to $N^2$ as a result of  self-phase-locking  of the beams due to two(three)-dimensional
DFB formed in a VFEL resonator. When  beams are generated, there
is always a certain  spread in the  times when the produced
electron bunches enter a VFEL resonator, which results from the
instability of the current generation in a diode gap.

In the present paper it is shown that despite this spread in the
entry times of the beams, a two(three)-dimensional feedback formed
in a VFEL resonator gives rise to the self-phase-locking process .
Equations are derived that describe the process of radiation
generation  by several electron bunches and enable studying the
process of superradiation produced by the bunches, depending on
the difference between the times of their entry into the
resonator. It is also shown that similar processes of
self-phase-locking, which lead to the power increase $W\sim N^2$
can also be observed when generation from electron bunches is
excited in a system of several relativistic BWOs, whose corrugated
waveguides form a single resonator by means of, e.g., slots (long
bridge corrugated waveguides or diffraction grating) made in a
cylinder along the axis of the waveguides.

\section{Lasing equations for
 the system with a photonic crystal (diffraction grating) with changing parameters}
\label{vp_sec:2}

In the general case the equations, which describe lasing process,
follow from the Maxwell equations:
\begin{eqnarray}
& \hspace{1cm}& \textrm{rot} \vec{H}=\frac{1}{c}\frac{\partial
\vec{D}}{\partial t}+\frac{4 \pi}{c} \vec{j}, ~\textrm{rot}
\vec{E}=-\frac{1}{c}\frac{\partial
\vec{H}}{\partial t}, \nonumber \\
&  &\textrm{div} \vec{D}=4 \pi \rho,~\frac{\partial \rho}{\partial
t}+ \textrm{div} \vec{j}=0, \label{vp_sys0}
\end{eqnarray}
here $\vec{E}$ and $\vec{H}$ are the electric and magnetic fields,
$\vec{j}$ and $\rho$ are the current and charge densities, the
electromagnetic induction $D_i(\vec{r},t^{\prime})=\int
\varepsilon_{il}
(\vec{r},t-t^{\prime})E_l(\vec{r},t^{\prime})dt^{\prime}$ and,
therefore, $D_i(\vec{r},\omega)=\varepsilon_{il}
(\vec{r},\omega)E_l(\vec{r},\omega)$, the indices $i,l=1,2,3$
correspond to the axes $x,y,z$, respectively.

%-----------------

The current and charge densities are respectively defined as:
\begin{equation}
\vec{j}(\vec{r},t)=e
\sum_{\alpha}\vec{v}_{\alpha}(t)\delta(\vec{r}-\vec{r}_{\alpha}(t)),~
\rho(\vec{r},t)=e
\sum_{\alpha}\delta(\vec{r}-\vec{r}_{\alpha}(t)), \label{eq2}
\end{equation}
where $e$ is the electron charge, $\vec{v}_{\alpha}$ is the
velocity of the particle $\alpha$ ($\alpha$ numerates the beam
particles),
\begin{equation}
 \frac{d
\vec{v}_{\alpha}}{dt}=\frac{e}{m \gamma_{\alpha}}\left\{
 \vec{E}(\vec{r}_{\alpha}(t),t)+\frac{1}{c} [
\vec{v}_{\alpha}(t) \times \vec{H}(\vec{r}_{\alpha}(t),t)
 ]-
 \frac{\vec{v}_{\alpha}}{c^2}(\vec{v}_{\alpha}(t)\vec{E}(\vec{r}_{\alpha}(t),t))
\right\},\label{vp_sys1}
\end{equation}
here $\gamma_{\alpha}=(1-\frac{v_{\alpha}^2}{c^2})^{-\frac{1}{2}}$
is the Lorentz-factor, $\vec{E}(\vec{r}_{\alpha}(t),t)$
($\vec{H}(\vec{r}_{\alpha}(t),t)$) is the electric (magnetic)
field at the point of location $\vec{r}_{\alpha}$ of  particle
$\alpha$.

It should be recalled that  (\ref{vp_sys1}) can also be written as
\begin{equation}
 \frac{d
\vec{p}_{\alpha}}{dt}=m\frac{d \gamma_{\alpha}
v_{\alpha}}{dt}={e}\left\{
 \vec{E}(\vec{r}_{\alpha}(t),t)+\frac{1}{c} [
\vec{v}_{\alpha}(t) \times \vec{H}(\vec{r}_{\alpha}(t),t)
 ]
\right\},\label{vp_sys11}
\end{equation}
where $p_{\alpha}$ is the particle momentum.

Let us recall here that the change in the particle energy through
the its interaction with electromagnetic fields is described by
the equation
\begin{equation}
\label{en} mc^2\frac{d\gamma_{\alpha}}{dt}=e\vec v_{\alpha}\vec
E(\vec r_{\alpha}(t),t).
\end{equation}

%-------------
Combining the equations in (\ref{vp_sys0}), we obtain:
\begin{equation}
-\Delta
\vec{E}+\vec{\nabla}(\vec{\nabla}\vec{E})+\frac{1}{c^2}\frac{\partial^2
\vec{D}}{\partial t^2}=-\frac{4 \pi}{c^2} \frac{\partial
\vec{j}}{\partial t}. \label{vp_sys01}
\end{equation}

%.........................................

The dielectric permittivity tensor can be expressed as
$\hat{\varepsilon}(\vec{r})=1+\hat{\chi}(\vec{r})$, where
$\hat{\chi}(\vec{r})$ is the dielectric susceptibility.
When $\hat{\chi} \ll 1$,  (\ref{vp_sys01}) can be rewritten as:

\begin{equation}
\Delta \vec{E}(\vec{r},t)-\frac{1}{c^2}\frac{\partial^2}{\partial
t^2}
\int \hat{\varepsilon}(\vec{r},t-t^{\prime})
\vec{E}(\vec{r},t^{\prime}) dt^{\prime}
=4 \pi \left( \frac{1}{c^2} \frac{\partial
\vec{j}(\vec{r},t)}{\partial t} + \vec{\nabla} \rho (\vec{r},t)
\right). \label{vp_1}
\end{equation}

When the grating is ideal $\hat{\chi}(\vec{r}) = \sum_{\tau}
\hat{\chi}_{\tau} (\vec{r}) e^{i \vec{\tau} \vec{r}}$, where
$\vec{\tau}$ is the reciprocal lattice vector.

Let the photonic crystal (diffraction grating) period be smoothly
varied with distance, which is much greater then the diffraction
grating {(ptotonic crystal lattice)} period.
It is convenient in this case to present the susceptibility
$\hat{\chi}(\vec{r})$ in the form, typical of the theory of X-ray
diffraction in crystals with lattice distortion \cite{bar12}:
\begin{equation}
\hat{\chi}(\vec{r})=\sum_{\tau} e^{i \Phi_{\tau}(\vec{r})}
\hat{\chi}_{\tau} (\vec{r}), \label{vp_chi01}
\end{equation}
where $\Phi_{\tau}(\vec{r})=\int \vec{\tau} (\vec{r}^{\,\prime})
d\vec{l}^{\prime}$, $\vec{\tau} (\vec{r}^{\,\prime})$ is the
reciprocal lattice  vector in the vicinity of the point
$\vec{r}^{\,\prime}$.
In contrast to the theory of X-rays diffraction, in the case under
consideration $\hat{\chi}_{\tau}$ can also depend on $\vec{r}$.
Moreover, $\hat{\chi}_{\tau}$ depends on the volume of the lattice
unit cell $\Omega$, which can be significantly varied for
diffraction gratings (photonic crystals), as distinct from natural
crystals.
The volume of the unit cell $\Omega(\vec{r})$  depends on
coordinate and, for example, for a cubic lattice it is determined
as
$\Omega(\vec{r})=\frac{1}{d_1(\vec{r})d_2(\vec{r})d_3(\vec{r})}$,
where $d_i$ are the lattice periods.
If $\hat{\chi}_{\tau} (\vec{r})$ does not depend on $\vec{r}$, the
expression (\ref{vp_chi01}) converts to that usually used for
X-rays in crystals with lattice distortion.

Recall here that  for an ideal crystal without lattice
distortions, the wave, which propagates in the crystal can be
presented as a superposition of plane waves:
\begin{equation}
\vec{E}(\vec{r},t) = \sum_{\vec{\tau}=0}^{\infty}
\vec{A}_{\vec{\tau}} e^{i(\vec{k}_{\tau} \vec{r}-\omega t)},
\label{vp_field}
\end{equation}
where $\vec{k}_{\tau}=\vec{k}+\vec{\tau}$.

Let us now use the fact that in the case under consideration the
typical length for the change of the lattice parameters
significantly exceeds the lattice period. Then the field inside
the crystal with lattice distortion can be expressed similarly to
(\ref{vp_field}), but with $\vec{A}_{\vec{\tau}}$ depending on
$\vec{r}$  and $t$ and changing noticeably at the distances much
greater than the lattice period.

Similarly, the wave vector should be considered as a slowly
changing function of a coordinate.

According to the above, let us find the solution of (\ref{vp_1})
in the form:
\begin{equation}
\vec{E}(\vec{r},t)=\textrm{Re} \left\{
\sum_{\vec{\tau}=0}^{\infty}  \vec{A}_{\vec{\tau}}
e^{i(\phi_{\tau} (\vec{r})-\omega t)} \right\}, \label{vp_*}
\end{equation}
where $\phi_{\tau} (\vec{r})= \int_0^{\vec{r}} k(\vec{r}) d
\vec{l}+ \Phi_{\tau}(\vec{r})$, where $k(\vec{r})$ can be found as
a solution of the dispersion equation in the vicinity of the point
with the coordinate vector $\vec{r}$, integration is made over the
quasiclassical trajectory, which describes motion of the
wavepacket in the crystal with lattice distortion.

Now let us consider the case when all the waves participating in
the diffraction process lie in a plane (coupled wave diffraction,
multiple-wave diffraction), i.e., all the reciprocal lattice
vectors $\vec{\tau}$ lie in one plane.
Suppose the wave polarization vector is orthogonal to the plane of
diffraction.

Let us rewrite (\ref{vp_*}) in the form
\begin{equation}
\vec{E}(\vec{r},t)=\vec{e}\,E(\vec{r},t)=\vec{e} \, \textrm{Re}
\left\{
 {A}_1
e^{i(\phi_{1} (\vec{r})-\omega t)}+ {A}_2 e^{i(\phi_{2}
(\vec{r})-\omega t)} + ... \right\}, \label{vp_*1}
\end{equation}
where
\begin{equation}
\phi_1(\vec{r})=\int_0^{\vec{r}} \vec{k}_1(\vec{r}^{\, \prime}) d
\vec{l}, \label{vp_phi1}
\end{equation}
\begin{equation}
\phi_2(\vec{r})=\int_0^{\vec{r}} \vec{k}_1(\vec{r}^{\,\prime}) d
\vec{l} + \int_0^{\vec{r}} \vec{\tau}(\vec{r}^{\,\prime}) d
\vec{l}. \label{vp_phi2}
\end{equation}

Then multiplying (\ref{vp_1}) by $\vec{e}$, one can get:
\begin{equation}
\Delta {E}(\vec{r},t)-\frac{1}{c^2}\frac{\partial^2}{\partial t^2}
\int
\hat{\varepsilon}(\vec{r},t-t^{\prime}){E}(\vec{r},t^{\prime})
dt^{\prime}
=4 \pi \vec{e} \left( \frac{1}{c^2} \frac{\partial
\vec{j}(\vec{r},t)}{\partial t} + \vec{\nabla} \rho (\vec{r},t)
\right). \label{vp_2}
\end{equation}
Applying the equality $\Delta {E}(\vec{r},t)=\vec{\nabla}
(\vec{\nabla} E)$ and using (\ref{vp_*1}), we obtain
\begin{eqnarray}
\hspace{-1 cm} \Delta (A_1 e^{i(\phi_{1} (\vec{r})-\omega t)})=
e^{i(\phi_{1} (\vec{r})-\omega t)}
[2i  \vec{\nabla} \phi_1 \vec{\nabla} A_1 +i \vec{\nabla}
\vec{k}_1 (\vec{r}) A_1  - k_1^2(\vec{r})  A_1], \label{vp_3a}
\end{eqnarray}

Therefore, substitution of the above expression into (\ref{vp_2})
gives the following system:
\begin{eqnarray}
& & \frac{1}{2} e^{i(\phi_{1} (\vec{r})-\omega t)} \left[ 2i
\vec{k}_1(\vec{r}) \vec{\nabla} A_1 +i \vec{\nabla}
\vec{k}_1 (\vec{r}) A_1  - k_1^2(\vec{r}) A_1  \right.\nonumber \\
& & \left.+ \frac{\omega^2}{c^2} \varepsilon_0(\omega,\vec{r}) A_1
+ i \frac{1}{c^2} \frac{\partial \omega^2
\varepsilon_0(\omega,\vec{r})}{\partial \omega} \frac{\partial
A_1}{\partial t} + \frac{\omega^2}{c^2}
\varepsilon_{-\tau}(\omega,\vec{r}) A_2 \right.\nonumber\\
& &+ \left.i \frac{1}{c^2} \frac{\partial \omega^2
\varepsilon_{-\tau}(\omega,\vec{r})}{\partial \omega}
\frac{\partial A_2}{\partial t}
\right] \nonumber \\
& & + \textrm{~conjugated~terms~}
=4 \pi \vec{e} \left( \frac{1}{c^2} \frac{\partial
\vec{j}(\vec{r},t)}{\partial t} + \vec{\nabla} \rho (\vec{r},t)
\right), \nonumber \\
& & \frac{1}{2} e^{i(\phi_{2} (\vec{r})-\omega t)} [ 2i
\vec{k}_2(\vec{r}) \vec{\nabla} A_2 +i \vec{\nabla}
\vec{k}_2 (\vec{r}) A_2  - k_2^2(\vec{r}) A_2  \nonumber \\
& & + \frac{\omega^2}{c^2} \varepsilon_0(\omega,\vec{r}) A_2 + i
\frac{1}{c^2} \frac{\partial \omega^2
\varepsilon_0(\omega,\vec{r})}{\partial \omega} \frac{\partial
A_2}{\partial t} + \frac{\omega^2}{c^2}
\varepsilon_{\tau}(\omega,\vec{r}) A_1 \nonumber\\
& &+ i \frac{1}{c^2} \frac{\partial \omega^2
\varepsilon_{\tau}(\omega,\vec{r})}{\partial \omega}
\frac{\partial A_1}{\partial t}
] \nonumber \\
& & + \textrm{~conjugated~terms~}
=4 \pi \vec{e} \left( \frac{1}{c^2} \frac{\partial
\vec{j}(\vec{r},t)}{\partial t} + \vec{\nabla} \rho (\vec{r},t)
\right),  \label{vp_3}
\end{eqnarray}
where vector $\vec{k}_2 (\vec{r})=\vec{k}_1 (\vec{r})+
\vec{\tau}$,
$\varepsilon_0(\omega,\vec{r})=1 + {\chi}_{0}(\vec{r})$, here the
notation ${\chi}_{0} (\vec{r})={\chi}_{\tau=0} (\vec{r})$ is used,
$\varepsilon_{\tau}(\omega,\vec{r})={\chi}_{\tau} (\vec{r})$.
Note here that for a numerical analysis of (\ref{vp_3}), if
${\chi}_{0} \ll 0$, it is convenient to take  vector $\vec{k}_1
(\vec{r})$ in the form $\vec{k}_1
(\vec{r})=\vec{n}\sqrt{k^2+\frac{\omega^2}{c^2} \chi_0(\vec{r})}$.

Let us multiply the first equation by $e^{-i(\phi_{1}(\vec
r)-\omega t)}$ and the second by $e^{-i(\phi_{2}(\vec r)-\omega
t)}$.
This procedure enables neglecting the conjugated terms, which
appear fast oscillating (when averaging over the oscillation
period they become zero).

Considering the right-hand side of (\ref{vp_3}), let us take into
account that microscopic currents and densities are the sums of
terms, containing delta-functions, therefore, the right-hand side
can be rewritten as:
\begin{eqnarray}
\label{fun} & & e^{-i(\phi_{1}(\vec r)-\omega t)}4 \pi \vec{e}
\left( \frac{1}{c^2} \frac{\partial \vec{j}(\vec{r},t)}{\partial
t} + \vec{\nabla} \rho (\vec{r},t)
\right) \\
& &= - \frac{4 \pi i \omega e}{c^2}  \vec{e} \sum_{\alpha} \vec
{v}_{\alpha} (t) \delta(\vec r -\vec r_{\alpha}(t))
e^{-i(\phi_{1}(\vec r)-\omega t)} \, \theta (t-t_{\alpha}) \,
\theta (T_{\alpha}-t). \nonumber
\end{eqnarray}

Here $t_\alpha$ is the time of entrance of particle $\alpha$  to
the resonator, $T_\alpha$ is the time of particle leaving the
resonator, $\theta-$functions in (\ref{fun}) indicate that for the
time moments preceding $t_{\alpha}$ and following $T_{\alpha}$,
the particle  ${\alpha}$ does not contribute to the process.

Upon averaging the system of equations over the oscillation period
$\frac{2\pi}{\omega}$, we can write:
\begin{eqnarray}
\label{5mod} & &\left[2i \vec{k}_1(\vec{r}) \vec{\nabla} A_1 +i
\vec{\nabla} \vec{k}_1 (\vec{r}) A_1  - k_1^2(\vec{r}) A_1 +
\frac{\omega^2}{c^2} \varepsilon_0(\omega,\vec{r}) A_1
\right.\nonumber
\\
& & \left. + i \frac{1}{c^2} \frac{\partial \omega^2
\varepsilon_0(\omega,\vec{r})}{\partial \omega} \frac{\partial
A_1}{\partial t} + \frac{\omega^2}{c^2}
\varepsilon_{-\tau}(\omega,\vec{r}) A_2 + i \frac{1}{c^2}
\frac{\partial \omega^2
\varepsilon_{-\tau}(\omega,\vec{r})}{\partial \omega}
\frac{\partial A_2}{\partial t}\right]
\\
 & &= - \frac{8\pi i\omega
e}{c^2}\sum_{\alpha}\int^{t+\frac{2\pi}{\omega}}_{t}\vec e\,\vec
v_{\alpha}(t')\delta(\vec r-\vec r_{\alpha}(t',t_{\alpha}, \vec
r_{\alpha 0})e^{-i\varphi_1[\vec r_{\alpha}(t',t_{\alpha},\vec
r_{\alpha\,
0}),t']}\theta(t'-t_{\alpha})\theta(t_{\alpha}-t'),\nonumber
\end{eqnarray}
where the phase $\varphi_1[\vec r_{\alpha}(t' ,t _{\alpha},\vec
r_{\alpha\, 0}),t']=\phi_1(\vec r_{\alpha}(t',t_{\alpha}, \vec
r_{\alpha\,0})-\omega t'$.
\begin{eqnarray}
\label{5mod2} & &\left[2i \vec{k}_2(\vec{r}) \vec{\nabla} A_2 +i
\vec{\nabla} \vec{k}_2 (\vec{r}) A_2  - k_2^2(\vec{r}) A_2 +
\frac{\omega^2}{c^2} \varepsilon_0(\omega,\vec{r}) A_2
\right.\nonumber
\\
& & \left. + i \frac{1}{c^2} \frac{\partial \omega^2
\varepsilon_0(\omega,\vec{r})}{\partial \omega} \frac{\partial
A_2}{\partial t} + \frac{\omega^2}{c^2}
\varepsilon_{\tau}(\omega,\vec{r}) A_1 + i \frac{1}{c^2}
\frac{\partial \omega^2
\varepsilon_{\tau}(\omega,\vec{r})}{\partial \omega}
\frac{\partial A_1}{\partial t}\right]
\\
 & &= -\frac{8\pi i\omega
e}{c^2}\sum_{\alpha}\int^{t+\frac{2\pi}{\omega}}_{t}\vec e\,\vec
v_{\alpha}(t')\delta(\vec r-\vec r_{\alpha}(t',t_{\alpha}, \vec
r_{\alpha 0})e^{-i\varphi_2[\vec r_{\alpha}(t',t_{\alpha},\vec
r_{\alpha\,
0}),t']}\theta(t'-t_{\alpha})\theta(t_{\alpha}-t'),\nonumber
\end{eqnarray}
where the phase $\varphi_2[\vec r_{\alpha}(t',t_{\alpha},\vec
r_{\alpha\, 0}),t']=\phi_2(\vec r_{\alpha}(t',t_{\alpha}, \vec
r_{\alpha\,0})-\omega t'$ and $\varepsilon_0=1+\chi_0$,
$\varepsilon_{\tau}=\chi_{\tau}$.

When several ($N$ number) electron beams move through a spatially
periodic medium, the sum $\sum_{\alpha}$ over the particles can be
represented as a sum of contributions coming from individual
electron beams to the total current:
\[
\sum_{\alpha}=\sum_{\alpha_1}+\sum_{\alpha_2}+...\sum_{\alpha\,N},
\]
where $N$ is the number of electron beams.

Using the definitions of $\phi_m$ (see (\ref{vp_phi1}),
(\ref{vp_phi2}) we can obtain the following relationship for the
phases $\varphi_m$:
\begin{equation}
\label{ph1}
\frac{d\varphi_m}{dt}=\vec k_m(\vec r_{\alpha}(t, t_{\alpha},
r_{\alpha\,0}))\vec v_{\alpha}(t)-\omega
\end{equation}
and
\begin{equation}
\label{ph2} \frac{d^2\varphi_m}{dt^2}=\vec
v_{\alpha}(t)\frac{d\vec k_m(r_{\alpha}(t))}{dt}+\vec
k_m(r_{\alpha}(t))\frac{d\vec v_{\alpha}}{dt}.
\end{equation}
Equations (\ref{vp_sys1})--(\ref{en}), describing particle motion
in electromagnetic fields, and equations
(\ref{5mod})--(\ref{5mod2}) for the fields are written using
slowly changing amplitudes $A$ and phases $\varphi_m=\phi_m-\omega
t$. They give a closed, nonlinear set of equations that defines
the amplitude $A_m$ and  phases $\varphi_m$ (as well as the change
in the energy of particles interacting with the fields) and can be
numerically analyzed using, say, the large-particle method.

Because of random distribution of particles in the bunches
incident on a resonator (electromagnetic, photonic crystal), the
times $t_{\alpha}$ of particle entry  into the resonator as well
as the distribution of the entry point coordinates $\vec r_s$ of
the bunch particles over the entire surface of the resonator are
random. Each  bunch  also has a certain distribution of initial
velocities $\vec v_{\alpha}^{(0)}$. This enables one to average
(\ref{5mod}) and (\ref{5mod2}) over the distribution of the
quantities $t_{\alpha}$$\vec r_s$, and $\vec v_{\alpha}^{(0)}$.
Such averaging can be made by generalizing the averaging method
developed for the case of one-dimensional generators like TWT,
BWO, FEL  to the case of a non-one-dimensional distributed
feedback (DFB) The equations obtained as a result of such
averaging in a stationary  case when one beam moves in a VFEL
resonator are given in \cite{bar12}.

We shall further consider self-phase-locking arising when photons
are emitted by several electron beams in a spatially periodic VFEL
resonator in the case of quasi-Cherenkov (diffraction) spontaneous
radiation mechanism (recall here that this radiation  mechanism
underlies the operation of conventional one-dimensional TWTs and
BWOs).

For better understanding, let us suppose now that a strong
magnetic field is applied for beam guiding through the generation
area. Electron beams move along the direction of this field. Let
us choose the direction of beam motion as the $z$-axis. We shall
also consider the case when the period of the resonator's
diffraction grating changes along the direction of the $z$-axis.
In this case, equations (\ref{5mod}) and (\ref{5mod2}) can be
presented in the form:
\begin{eqnarray}
\label{ins1}
& & \left[2i \vec{k}_1(\vec{r}) \vec{\nabla} A_1 +i
\vec{\nabla} \vec{k}_1 (\vec{r}) A_1  - k_1^2(\vec{r}) A_1
\frac{\omega^2}{c^2} \varepsilon_0(\omega,\vec{r}) A_1
\right.\nonumber
\\
& & \left. + i \frac{1}{c^2} \frac{\partial \omega^2
\varepsilon_0(\omega,\vec{r})}{\partial \omega} \frac{\partial
A_1}{\partial t} + \frac{\omega^2}{c^2}
\varepsilon_{-\tau}(\omega,\vec{r}) A_2 + i \frac{1}{c^2}
\frac{\partial \omega^2
\varepsilon_{-\tau}(\omega,\vec{r})}{\partial \omega}
\frac{\partial A_2}{\partial t}\right]\nonumber
\\
& & =-\frac{8\pi i \omega e\vartheta_1}{c^2}g_1(r_{\perp}, z, t),
\end{eqnarray}
where
\begin{eqnarray*}
g_1(r_{\perp}, z, t) &= &
\sum_{N}\langle\langle\int^{t+\frac{2\pi}{\omega}}_{t}
\sum_{\alpha_N}u_{\alpha_N}(t)\delta(r_{\perp}-r_{\alpha_N\perp})\delta(z-z_{\alpha_N}(t',t_{\alpha_N},
u_{\alpha_N}^{(0)}))
\\
&  \times & e^{-i\vec k_{\perp}\vec
r_{\alpha_N\perp}}e^{-i[\phi_1(z_{\alpha}(t',t_{\alpha_N},u_{\alpha_N}^{(0)}))-\omega
t']} \theta(t'-t_{\alpha}) \theta(T_{\alpha-t'})\rangle\rangle d
t',
\end{eqnarray*}
\begin{eqnarray}
\label{ins2} & &\left[2i \vec{k}_2(\vec{r}) \vec{\nabla} A_2 +i
\vec{\nabla} \vec{k}_2 (\vec{r}) A_2  - k_2^2(\vec{r}) A_2 +
\frac{\omega^2}{c^2} \varepsilon_0(\omega,\vec{r}) A_2
\right.\nonumber
\\
& & \left. + i \frac{1}{c^2} \frac{\partial \omega^2
\varepsilon_0(\omega,\vec{r})}{\partial \omega} \frac{\partial
A_2}{\partial t} + \frac{\omega^2}{c^2}
\varepsilon_{\tau}(\omega,\vec{r}) A_1 + i \frac{1}{c^2}
\frac{\partial \omega^2
\varepsilon_{\tau}(\omega,\vec{r})}{\partial \omega}
\frac{\partial A_1}{\partial t}\right]\nonumber
\\
& &=-\frac{8\pi i \omega e\vartheta_2}{c^2}g_2(r_{\perp}, z, t),
\end{eqnarray}
where
\begin{eqnarray*}
g_2(r_{\perp}, z, t) &= &
\sum_{N}\langle\langle\int^{t+\frac{2\pi}{\omega}}_{t}
\sum_{\alpha_N}u_{\alpha_N}(t)\delta(r_{\perp}-r_{\alpha_N\perp})\delta(z-z_{\alpha_N}(t',t_{\alpha_N},
u_{\alpha_N}^{(0)}))
\\
&  \times & e^{-i\vec k_{\perp}\vec
r_{\alpha_N\perp}}e^{-i[\phi_2(z_{\alpha}(t',t_{\alpha_N},u_{\alpha_N}^{(0)}))-\omega
t']} \theta(t'-t_{\alpha}) \theta(T_{\alpha-t'})\rangle\rangle d
t'.
\end{eqnarray*}
Here $\phi_1(z_{\alpha}(t',t_{\alpha_N},
u_{\alpha_N}^{(0)}))=\int_0^{z_{\alpha}(t',t_{\alpha_N},
u_{\alpha_N}^{(0)}))}k_{1z}(z')dz'$ and
$\phi_2(z_{\alpha}(t',t_{\alpha_N},
u_{\alpha_N}^{(0)}))=\int_0^{z_{\alpha}(t',t_{\alpha_N},
u_{\alpha_N}^{(0)}))}k_{2z}(z')dz'$. If the period of the
diffraction grating is constant along the $z$-axis, then
$\phi_1=k_1z_{\alpha}(t', t_{\alpha_N}, u_{\alpha_N}^{(0)})$. The
sign $\langle\langle...\rangle\rangle$ means averaging over the
distribution of beam particles over the transverse coordinate of
the entry points, the entry time, and over the velocity
distribution of the beams entering the resonator;
\[
\vartheta_m=\sqrt{1-\frac{\omega^2}{\beta^2 k_m^2 c^2}},\quad \beta^2=1-\frac{1}{\gamma^2},
\quad \vec k_1=\vec k_{\tau=0}, \quad \vec k_2=\vec k_1+\vec{\tau}.
\]

Let $\rho(\vec r_{\perp})$ denote the particle density in the plane transverse relative to the particle velocity and $\dot{n}$
denote the number of particles traversing the  inner surface of the resonator per unit time.
We also make use of the fact that
\[
\delta(z-z_{\alpha_N}(t',t_{\alpha_N},
u_{\alpha_N}^{(0)}))=\delta(t'-\tau(z,t_{\alpha_N},
u_{\alpha_N}^{(0)}))/\left|\frac{\partial
z_{\alpha_N}(t',t_{\alpha_N}, u_{\alpha_N}^{(0)})}{\partial
t'}\right|,
\]
where
\[
\tau(z,t_{\alpha_N}, u_{\alpha_N}^{(0)})=
t_{\alpha_N}+\int_0^z\frac{d
z'}{u_N(z',t_{\alpha_N}u_{\alpha_N}^{(0)})}.
\]
Here $\tau $ is the time required for the particle entering the
interaction area $z=0$ at time $t_{\alpha}$ at initial  speed
$u_{\alpha_N}^{(0)}$ to reach the point $z$ and
$u_N(z,t_{\alpha_N},u_{\alpha_N}^{(0)})$ is the speed at point $z$
for the particle whose initial speed  at $z=0$ and time
$t_{\alpha_N}$ equaled $u_{\alpha_N}^{(0)}$, while $\frac{\partial
z_{\alpha_N}(t',t_{\alpha_N}, u_{\alpha_N}^{(0)})}{\partial t'}$
is the particle speed at time $t$ if at $z=0$ and $t_{\alpha_N}$
its speed was $u_{\alpha_N}^{(0)}$. Obviously, for the expression
to be finite, this speed should not vanish; otherwise, such
transformation of the $\delta$-function becomes invalid (this is
possible, for example, in the case when the beam's current exceeds
the so-called limiting current and the virtual cathode is formed).

We shall further assume that  the particles are not retarded
significantly  during the interaction and write the quantity in
the denominator of $\frac{\partial z_{\alpha_N}}{\partial
t}=u_{\alpha_N}^{(0)}$. As a result, we have
$\frac{\dot{n}}{u_\alpha}=\rho_z(t_{\alpha})$, where
$\rho_z(t_{\alpha})$ is the  beam's time-dependent density
distribution along the $z$-axis.

Let us also suppose that the duration of the bunches injected into
the resonator is larger than the period $\frac{2\pi}{\omega}$ of
excited oscillations of the electromagnetic wave. Taking into
account the relationship for
$\tau=t_{\alpha_N}+\frac{z}{u_{\alpha}}$, which follows in this
case from the $\delta$-function, the distribution density
$\rho_z(t'_{\alpha})$ in this case  can be removed from the sign
of integration  over the  entry time $t_{\alpha_N}$ at point
$t'-\frac{z}{u_{\alpha}}$, and so we have
\[\rho_z(t_{\alpha})=\rho_z(t'-\frac{z}{u_{\alpha}}).
\]
In a real situation, the distribution of the beam's longitudinal
velocity is much less than the average longitudinal velocity of
the bunch, so we can remove the density $\rho_z$ from the sign of
integration describing averaging over the velocity distribution.
However, analyzing the phase dependence on the velocity
distribution of particles in a bunch, one should bear in mind that
the velocity distribution can appreciably affect the phase.

As a result, we can obtain the following expression for $g_n$:
\begin{eqnarray}
\label{eq_g} g_n(r_{\perp}, z, t) & = & \sum _{N}e^{-i\vec
k_{\perp}\vec r_{\perp}}\rho_N(r_{\perp}, z- u_N t)\nonumber
\\
& \times & u_N
\int_{t-\frac{z}{u_N}}^{t-\frac{z}{u_N}+\frac{2\pi}{\omega}}e^{-i[\phi_n(z(\tau,
t_0, u^{(0)}))-\omega\tau(t_{\alpha},u^{(0)})]}
f(u_N^{(0)})du_N^{(0)}dt_0.
\end{eqnarray}
Here $d t_0$ means integration over the initial times,
$f(u^{(0)}_N)$ is the velocity distribution function in bunch $N$,
and $u_N$ is the average velocity of the $N$-th bunch; the
right-hand part of equation differs from zero at times $t>0$ from
the initial moment defined as the instant of time when  the first
particle of the first bunch enters the resonator.

The derived set of equations enables describing the process of
radiation from several beams in a spatially periodic system
(photonic crystal), including the case when the beams move
opposite to one another. The geometry when the beams move opposite
to one another can be used for beam diagnosing in the bunch-bunch
collision region in colliding-beam storage rings. For short
bunches, this set of equations describes the phenomenon of
super-radiance produced when several bunches of relativistic
particles pass through a VFEL resonator. Particularly, it is
possible to investigate radiation as a function of the difference
between the times of electron bunches entry into the photonic
crystal and as a function of the transverse distance between the
bunches moving in the electromagnetic (photonic) crystal.

According to the equations derived here, the electromagnetic field
induced in the crystal by different beams does not contain random
phases $r_{\alpha_N}$ and $t_{\alpha}$ any longer, and the total
field, as a result, is a coherent sum of the induced fields, which
means that the radiation power increases as ${W\sim E^2=(\sum_N
E_n)^2\simeq N^2 E_1}$. It should be noted that when the
parameters $\chi_{\tau}$ grow, $|\chi_{\tau}|\geq 1$, the
plane-wave expansion of the solution to Maxwell'equations, which
is used in the dynamical diffraction of waves in crystals,
requires that for accurate description of the radiation generation
process a larger number of waves should be considered. However, in
this case one can expand the electromagnetic field into the
analogue of the Wannier functions, which are used for describing
the band structure of electrons in crystals in the case of tight
binding.

Let us consider the following example.  Let a spatially periodic
resonator be formed by  axially corrugated cylindrical waveguides.
we shall choose the direction of the waguides' axis as the
$z$-axis. Coupled through the slots in their walls (long bridge
corrugated waveguides or diffraction grating), the waveguide form
a single spatially periodic electrodynamical system. In the
general case, we have a 2D periodic system in the ($x,y$) plane,
orthogonal to the $z$-axis. The beams move along the $z$-axis.
 Depending on the position of the
cylinders in the transverse plane, square gratings or more
complicated structures  can be formed, e.g the cylinders can be
arranged in a circle (as it occurs in the magnetron).

To describe generation in this system, it is convenient start with
the expansion Maxwell's equations in terms of the eigenfuctions
$\vec Y_n(\vec r_{\perp})$ of this transverse grating (see a
similar approach used for describing the motion of fast electrons
in crystals in channeling regime (mode)) \cite{barbook}.
In this case the eigenfunction $\vec Y_n(\vec r_{\perp})$ is a sum
of the localized Wannier functions $W_n$
\[
\vec Y_{n\vec \kappa}(\vec r)=\sum_m W_n (\vec r_{\perp}-\vec
R_{\perp m})e^{i\vec \kappa\vec R_{\perp m}},
\]
where $\vec \kappa$ is the reduced wave vector, $n$ is the set of
indices defining stationary wave functions , e.g., the wave
function used for the formation of the structures periodic in the
transverse plane, and $\vec R_m$ is the coordinate of the centre
of the elementary cell $m$ of the structure periodic in the
transverse plane.

As a result, for the analysis of generation of radiation we obtain
one dimensional along the $z$-axis equations, where the excitation
current is the total current $I$ produced by the beams moving in
the system. Let us average the current $I$ over the electron entry
times, the distribution of the initial velocities,   and  the
distribution of electrons in the transverse plane (which in this
case have the peaks in the regions where the electrons from each
beam producing the current $I$ move).

As a result we obtain the equations similar in form to those used
for the analysis of the generation process induced by one beam
moving in a waveguide that is spatially periodic along the
$z$-axis axis (formed by , e.g., a corrugated waveguide of a
relativistic BWO.)
Hence, we can conclude that the considered  system,  excited by
several beams, generates common coherent radiation. Now, let us
give a more detailed consideration of the case when the resonator
is formed by the two elements of the grating.

When the resonator period is formed by corrugation of the
waveguide surface, we obtain a system consisting of two corrugated
waveguides coupled through, say,  a slot. For the BWO in the
stationary case when two stationary electron  beams move through
circular waveguides, this system was analyzed neglecting the
influence of  the  wave moving in the same direction \cite{balak}.
Using numerical analysis, the authors of \cite{balak} showed that
at certain parameters, a single-frequency oscillation mode is stet
in the system, i.e., in fact, coherent summation of the amplitudes
of the fields induced by two separate beams is possible. As
follows from the above analysis, such coherent summation is also
possible in the case non-stationary excitation of the system by
two pulses of electron beams.

Note here that the equations derived in this paper enable taking
account of the influence of a coherent wave on the generation
process in a system of several BWOs. Moreover, according to
\cite{bar12}, just in the range of parameters where the amplitudes
of the incident and diffracted waves are comparable, in a
two-three dimensional periodic system the increment of radiative
instability increases sharply and the threshold for the generation
start drops dramatically.

When applied to this case, general equations for describing the
excitation of two relativistic BWOs by two pulses of electron
beams can be written in the form:

Neglecting dispersion in considering the generation process in a
system of two BWOs with a constant grating period (corrugation
period), one can write these equations in the form:
\begin{eqnarray}\label{2bwt1}
\left\{\begin{array}{c}
   2ik_{1 z}\frac{\partial
A^a_1}{\partial z}  +  2 i\frac{\omega}{c^2}\frac{\partial
A^a_1}{\partial
t}+\left[\frac{\omega^2}{c^2}\varepsilon_0-k^2_{mn}-k^2_{1z}\right]
A_1^a\\
+\frac{\omega^2}{c^2}\varepsilon_{\tau}A_2^a+\frac{\omega^2}{c^2}\chi_{ab}^1
A^b_1+\frac{\omega^2}{c^2}\chi^2_{ab}A^b_2=\vec Y_{mn}\vec
j_1,\\
\\
 2ik_{2 z}\frac{\partial A^a_2}{\partial z}  +  2
i\frac{\omega}{c^2}\frac{\partial A^a_2}{\partial
t}+\left[\frac{\omega^2}{c^2}\varepsilon_0-k^2_{mn}-k^2_{2z}\right]
A_2^a \\
 +\frac{\omega^2}{c^2}\varepsilon_{\tau}A_2^a+\frac{\omega^2}{c^2}\chi_{ab}^1
A^b_1+\frac{\omega^2}{c^2}\chi^2_{ab}A^b_2=\vec Y_{mn}\vec
j_{1\tau},
\end{array}\right.
\end{eqnarray}

\begin{eqnarray}\label{2bwt2}
\left\{\begin{array}{c}
   2ik_{1 z}\frac{\partial
A^b_1}{\partial z}  +  2 i\frac{\omega}{c^2}\frac{\partial
A^b_1}{\partial
t}+\left[\frac{\omega^2}{c^2}\varepsilon_0-k^2_{mn}-k^2_{1z}\right]
A_1^b\\
+\frac{\omega^2}{c^2}\varepsilon_{\tau}A_2^b+\frac{\omega^2}{c^2}\chi_{ab}^1
A^b_1+\frac{\omega^2}{c^2}\chi^2_{ab}A^b_2=\vec Y_{mn}\vec
j_2,\\
\\
 2ik_{2 z}\frac{\partial A^b_2}{\partial z}  +  2
i\frac{\omega}{c^2}\frac{\partial A^b_2}{\partial
t}+\left[\frac{\omega^2}{c^2}\varepsilon_0-k^2_{mn}-k^2_{2z}\right]
A_2^b \\
 +\frac{\omega^2}{c^2}\varepsilon_{\tau}A_2^b+\frac{\omega^2}{c^2}\chi_{ab}^1
A^b_1+\frac{\omega^2}{c^2}\chi^2_{ab}A^b_2=\vec Y_{mn}\vec
j_{2\tau},
\end{array}\right.
\end{eqnarray}

We have for the  BWO mode
\[
A_1^{a(b)}(z=0)=0,\qquad A_2^{a(b)}(z=L)=0,
\]
where $L$ is the resonator length.

 For the case when more  than
two waveguides are involved, say, $N$ number - $N$ number of pairs
of equation sets are required, instead of  the two sets given
above, and the terms describing the waves produced by other
waveguides that are similar to $\chi_{ab}^1 A_1^b$ and
$\chi_{ab}^2 A_2^b$ in (\ref{2bwt1}), (\ref{2bwt2})  should be
added to each pair of equations.

It is worth noting that in a real case of arbitrary $\chi_{\tau}$,
the coefficients appearing in these equations should be considered
as phenomenological coefficients determined from the experiment on
the passage of an electromagnetic wave through such structures.

The derived system of equations enables describing the process of
generation excited by the combined pulses of electron bunches in a
periodic system of coupled periodic waveguides (artificial
electromagnetic crystal, VFEL resonator), which forms
self-phase-locking of coherent oscillations. The derived set of
equations enabled studying the dependence of the radiation power
on the difference between the times of the bunches'entry into the
resonator of such a periodic system. This equation set is
applicable to describing radiation produced by bunches with
various duration, and consequently in the case of  short bunches
it allows one to describe the phenomenon of superradiation and
phase-locking  in the system of several relativistic BWOs coupled
into the grating.

\section{Conclusion}

The above analysis shows that when a spatially periodic system of
a VFEL resonator is excited by several ($N$) pulsed electron beams
that enter the resonator at different times, the
two-(three)-dimensional distributed feedback, formed in the
resonator, gives rise to self-phase-locking  of the radiation
process and coherent collective oscillations, which result into a
square in $N$ increase of the radiation power $W$ with growing
number of beams: $W\sim N^2$. Such self-phase locking occurring in
a VFEL resonator makes it possible for us to consider the VFEL as
the self-phase-locking system. Using this equation set, one can
describe generation of generation  by several beams in different
modes: as superradiation from several electron beams and as
radiation from long beams. These equations also make it possible
to study the process of generation of superadiation as a function
of the difference between the times of bunches' entry into the
resonator.

\end{document}